          [1999/12/02 1.01 (PWD)]
\documentclass[letterpaper,twocolumn]{esapub} 

\usepackage{natbib}
\usepackage{epsfig}

\title{Flat-Fielding BATSE Occultation Data for use in a Hard X-Ray All Sky Survey}
\author[1]{S.E. Shaw} 
\author[1]{A.J. Bird} 
\author[1]{A.J. Dean} 
\author[1]{N. Diallo} 
\author[1]{C. Ferguson} 
\author[2]{J. Kn\"odlseder}
\author[1]{J.J. Lockley}
\author[1]{\\M.J. Westmore} 
\author[1]{D.R. Willis}
\affil[1]{Southampton University, UK}
\affil[2]{Centre d'Etude Spatiale des Rayonnements, France}

\begin{document}

\keywords{BATSE; Hard X-Rays; Survey}

\maketitle

\begin{abstract}
The BATSE mission aboard CGRO can be used to observe hard X-ray
sources by using the Earth occultation method.  This method relies on
measuring a step in the count rate profile in each BATSE detector as a
source rises above or sets below the Earth's limb.  A major problem in
determining the step sizes (and hence the flux) is in extracting the
steps from the varying background.  A technique for
flat-fielding the response of $\gamma$-ray detectors has been developed at Southampton.  The technique uses a dynamic Monte-Carlo model to simulate the dominant components of the gamma-ray background encountered by the
experiment at any point in its orbit.  A maximum likelihood imaging method is also being developed that will be used to make a sky survey with all $\sim$9 years of the BATSE CONTINUOUS data set in the 20 - 500 keV range.
An all sky map of 25 - 35 keV emission has been made using 60 days of data and has a 3$\sigma$ flux sensitivity of 14 mCrab. 
\end{abstract}

\section{Introduction}
The Burst and Transient Source Experiment (BATSE) consisted of eight
2025 cm$^{2}$ NaI(Tl) Large Area scintillation Detectors (LADs), sensitive from 20 keV -
2 MeV.  BATSE monitored the whole sky continuously from April 1991 -
June 2000 as an experiment aboard the Compton Gamma-Ray Observatory (CGRO).  The continuous whole sky
exposure makes BATSE an ideal facility for the production of the first
all sky hard X-ray survey since HEAO A-4 of 1977-1978
(\citet{heaoa4}).  A survey performed with BATSE will not
only be the first in the few hundreds of keV band but will
have the ability
to detect weak persistent sources and continuously monitor variable and transient sources.  The BATSE CONTINUOUS
data set contains usable count rates in 15 energy channels recorded in 2.048
s time bins.
\begin{figure*}[ht]
\centering
\epsfig{file=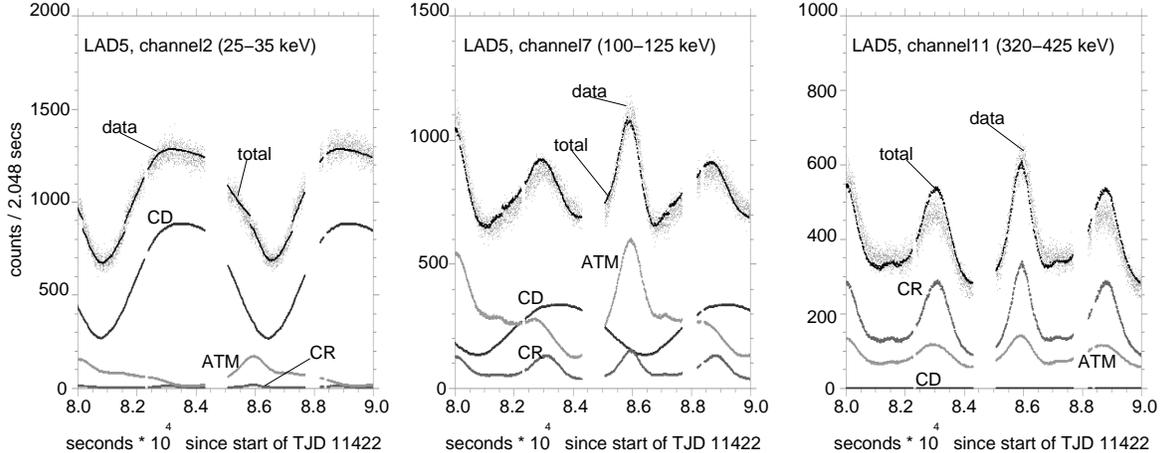, width=0.95\linewidth}
\caption{Comparison of the count rate profiles recorded by the BATSE
LAD 5 (dots) and the Southampton modelled data.  The total is the sum of the
cosmic diffuse emission (CD), the atmospheric albedo (ATM) and prompt
cosmic ray interactions (CR).} \label{fig:bgcomp}
\end{figure*}
\section{BATSE Earth Occultation Observations}
BATSE was developed primarily to detect and locate $\gamma$-ray bursts.
The individual BATSE detectors have no positional sensitivity, hence
to locate the position of a source of $\gamma$-rays the relative count
rates from several detectors are combined, along with the point spread function of each detector (\citet{pendleton99}).  Prior to launch
however it was realised that that the instrument could also be used as a
sensitive all sky monitor of point like $\gamma$-ray sources by using the Earth to modulate the flux at the
CGRO orbital period (\citet{paciesas85}).  Sharp steps are seen in the count rate profile of
the detectors as point sources rise and set over the Earth's limb,
the size of which give the source flux (\citet{batse2000}).  About 70\% of the whole sky is occulted twice per 90
minute orbit.  The precession of the orbit is such that the whole sky
undergoes occultation every 53 days.  The key to measuring the flux values correctly is in
accurately determining the varying background count rate.  \citet{ling2000} have recently published a catalogue of
34 moderately strong $\gamma$-ray sources observed with BATSE during the
first three phases of the CGRO mission (May 1991 through October
1994).  They use an empirical fit to find the background, which is
hampered by systematic errors and large computation time
(\citet{batse2000}), and make no attempt to image the sky.
\section{Southampton BATSE Background Model}
The BATSE model is based on The INTEGRAL Mass Model (TIMM), which is a GEANT based Monte-Carlo simulation
code devloped at Southampton to investigate the performance of the
instruments aboard the INTEGRAL mission pre-launch (\citet{TIMM}).  It
considers three components to the background; cosmic rays, cosmic diffuse gamma
rays and $\gamma$-rays from interactions of cosmic rays in the Earth's
atmosphere.  These same techniques are easily extended to other
experiments and has been proposed for use with the BAT burst monitor aboard SWIFT.
\subsection{Dynamic Generation of Background Components}
With BATSE we are interested in the dynamical behaviour of the
background within the CGRO orbit.  The background is a complicated
function of the pointing of the detectors, the position of CGRO in the
Earth's magnetic field and its irradiation history.  A novel two step
method has been developed to account for the background at any point
in the CGRO lifetime.  Firstly CGRO is assumed to be in deep space
without any interference from Earth.  Here it receives isotropic
fluxes of cosmic rays, cosmic diffuse $\gamma$-rays and atmospheric $\gamma$-rays and a database,
containing information about all the events interacting in the BATSE
modules is created.  The second stage filters the database with the
physical criterea corresponding to CGRO's position and pointing.  This
information is used to create the background for any orbital position
and pointing.  The component due to the atmospheric $\gamma$-rays is
found by scaling and sampling of the detector pointings and finally
the radio-activation of the detector materials is simulated.  The first stage is computationally very intensive, but
only needs to be done once for the whole lifetime of the experiment.
To compute the second stage of the background simulation requires only
a few hours for all of the eight BATSE modules for one day.  Figure~\ref{fig:bgcomp} shows the excellent agreement between the
modelled data and the count rate profile in the BATSE LADs for three
energy channels. 
\begin{figure*}[t]
\centering
\epsfig{file=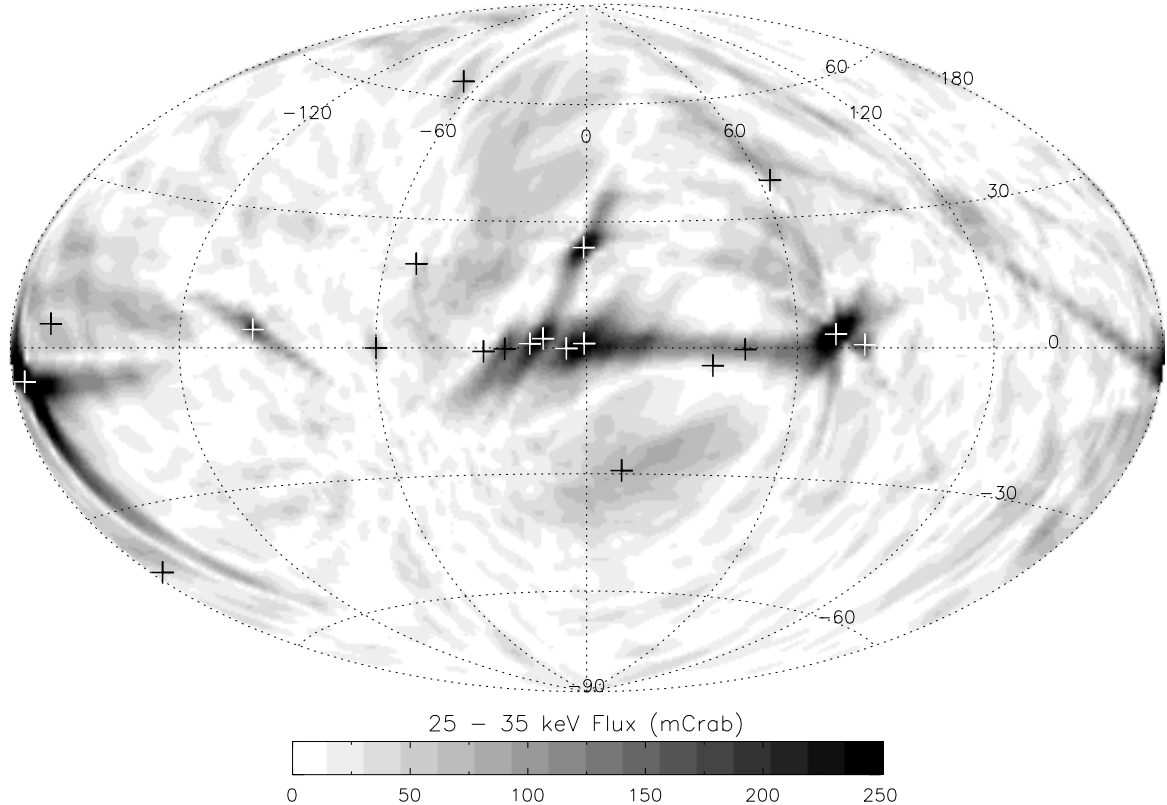, width=0.95\linewidth}
\caption{Map of the 25 - 35 keV flux from the
      whole sky for a total of 61 days from TJD 10840 to TJD 10900
      made using LIMBO.  The colour scale of the map
      has been limited to 250 mCrab (one quarter of the peak of 1 Crab) to show details at lower significance.  Crosses
      mark the actual positions of the objects given in table~\ref{tab:sources}.}\label{fig:map}
\end{figure*}
\begin{table}
  \begin{center}
    \caption{Approximate sensitivity and angular resolution of the imaging
    technique at $\sim$30 keV as a function of filter width, $f$.}
    \renewcommand{\arraystretch}{1.2}
    \begin{tabular}[h]{ccc}
      \hline
      $f$      & 3$\sigma$ flux     & Angular resolution \\
	(bins) & sensitivity (mCrab)& (HWHM$^{\circ}$)  \\
      \hline
      5   & 270 & 0.5 \\
      20  & 199 & 0.9 \\
      40  & 88 & 2.4 \\
      60  & 77 & 2.9 \\
      120 & 74 & 5.3 \\
      \hline \\
      \end{tabular}
    \label{tab:res}
  \end{center}
\end{table}
\section{Production of Sky Images}
Sky images are produced using a maximum likelihood method originally
developed by J. Kn\"odlseder at CESR (\citet{jurgen99}).  This method
has been adapted at Southampton for
use with the flat-fielded CONTINUOUS data and developed into the LIMBO code (Likelihood
Imaging Method for BATSE Occultation).
\subsection{Flat-Fielding of Data}
The data is flat-fielded by subtracting the modelled background and the
result passed to a simple differential filter of width $f$ bins, given by
\begin{equation}
d^{\prime}_{i} = \frac{\sum^{j=i + f}_{j=i}d_{j} - \sum^{j=i -
f}_{j=i}d_{j}}{f+1}.
\label{eqn:diff}
\end{equation}
The filtered data, $d^{\prime}_{i}$, is transformed from a count rate
containing occultation steps to a set of occultation peaks.  A square
step at bin $i$ will be transfromed to a triangular peak
of width $2f$.    The choice of $f$ has implications for the
sensitivity and angular resolution of the imaging technique; larger
$f$ will give a more significant occultation peak and improve
sensitivity, smaller $f$ returns less significant but sharper peaks
that will lead to better angular resolution.  The gain in sensitivity
from increasing $f$ begins to level out as the steps from other
sources enter the filter.  Table~\ref{tab:res}
summarises the sensitivity and angular resolution obtained from images made from
one day's observation of Cygnus X-1 using energy channel 2 ($\sim$25 -
35 keV).  The creation of images is discussed further in section~\ref{sec:image}.
The raw BATSE data contains occasional gaps where, for example, the
instrument is turned off during passage through the South Atlantic
Anomaly (SAA) or where contact is lost between CGRO and telemetry relay
satellites.  Left untreated the gaps cause large artificial peaks in the data
near each gap so a linear interpolation is used to fill the gaps in
the background subtracted data.   The gap data is not used further in the analysis.
\subsection{Maximum Likelihood Imaging}
\label{sec:image}
The imaging method used here differs from the Radon transform
method used by other groups (e.g. \citet{zhang93}) in that it produces
the results of statistical tests for emission at each point in
the map rather than a deconvolution of the data.  The occultation peak
dataset, $\vec{o}$, is fitted to a response vector, $\vec{e}$, which
describes the expected position, shape and amplitude of the peaks as a
function of source strength and position on the sky.  The fit
determines a scaling factor of the response vector, which is
proportional to the flux received from each point in the sky for all
eight LADs.  The scaling factor, $\alpha$, is given by
\begin{equation}
\alpha = \frac{\sum_{i}\frac{o_{i}e_{i}}{\sigma_{i}^{2}}}{\sum_{i}\frac{e_{i}^{2}}{\sigma_{i}^{2}}},
\label{eqn:alpha}
\end{equation}
where $\sigma_{i}$ is the statistical error on $o_{i}$.  A
maximum likelihood ratio test is used to determine the significance of
the flux detection and returns a skymap for a given grid of source
positions (\citet{cash76}).  The maximum likelihood ratio, $\lambda$, is distributed as
$\chi^{2}_{\nu}$ with $\nu$ degrees of freedom.  It is calculated
from the difference $C_{o} - C_{src}$, the $\chi^{2}$ statistics for
the null (no source is fitted to $\vec{o}$) and source (a source
with free strength is fitted to $\vec{o}$) hypotheses respectively,
where
\begin{eqnarray}
C_{o} & = & \sum_{i}\frac{o^{2}_{i}}{\sigma_{i}^{2}} ,\\
C_{src} & = & \sum_{i} \frac{(o_{i} - \alpha
e_{i})^{2}}{\sigma_{i}^{2}}. 
\label{eqn:csrc}
\end{eqnarray}
When searching for a known source $\nu = 1$ and hence the detection
significance, in gaussian $\sigma$, is given by $\surd \lambda$.  For
searches of new sources of $\gamma$-rays $\nu = 3$ (\citet{jurgen99}).
The image is built up from the superposition of the arcs caused by the
occultation of point sources by the Earth's limb.  The
angular resolution perpendicular to the limb depends on $\beta$, the angle
at the geocentre between the orbital plane and the source.  Sources
seen with $\beta\sim$45$^{\circ}$ yield good information since the
rising and setting limbs are almost orthogonal.
\par The maximum likelihood method has an advantage over the
Radon transform method in that there is no limit to the size of map that can
be generated.  The Radon transform is not well behaved for
regions much larger than $\sim$20$^{\circ}\times$20$^{\circ}$ (\citet{batse2000}).  Likelihood maps can be made for short periods of time and then added
together to increase source significance whilst retaining the ability
to investigate variability on short timescales.  The total
likelihood ratio from a total of individual daily maps, indexed by $d$, is 
$\lambda_{T} = \sum_{d} (\lambda_{d}-\Delta_{d})$ where
 \begin{equation}
\Delta_{d} = (\alpha_{T} -\alpha_{d})((\alpha_{T} - \alpha_{d})D_{d} - 2F_{d})
\end{equation}
and $F_{d}$ and $D_{d}$ are the daily values of the numerator and
denominator of equation~\ref{eqn:alpha} respectively.  The quantity
$\alpha_{T}$ is the total flux for all days being considered,
i.e. $\alpha_{T} =  \sum_{d}\alpha_{d} =\sum_{d}F_{d}/\sum_{d}D_{d}$.
\section{Results}
Figure~\ref{fig:map} shows the total likelihood map for the whole sky
created from channel 2 CONTINUOUS data for 60 days between Truncated
Julian Days (TJD) 10840 -
10900.  The energy calibration of BATSE channel 2 is such that the
range of photon energies it detects is approximately constant at
$\sim$9 keV with a lower limit of $\sim$25$\pm$5 keV.  The Crab, Cygnus
X-1 and several Galactic Centre sources have been identified along
with other faint sources as summarized in table~\ref{tab:sources}.
Some artifacts remain around the brighter sources, which we aim to
reduce by removing the brighter sources iteratively.  The 3$\sigma$ flux sensitivity of the map is 14 mCrab.  The map was
created using a 2$^{\circ}\times$2$^{\circ}$ sky grid and $f = 20$.
Table~\ref{tab:res} suggests that it would be worthwhile increasing
the grid resolution to 0.5$^{\circ}\times$0.5$^{\circ}$, which would
increase the computation time and disk space requirements by a factor
of 8.  The current maps are $\sim$600 kB in size per day and energy
channel and LIMBO takes $\sim$45
mins to run on a PentiumIII Linux workstation, although the code is not
fully optimised as yet.  On the basis of this preliminary analysis
LIMBO shows a lot of promise as a method of surveying the sky.  We
will continue to develop LIMBO with the aim of creating a database of
daily maps in each energy channel for the entire BATSE observation.
Sources found in the surveys will be
studied in more detail for location, variability and spectroscopy (see also presentations by Dean and Westmore at this meeting).
\begin{table}
  \begin{center}
    \caption{A selection of point sources of 25 - 35 keV $\gamma$-rays
    found from the sky map of figure~\ref{fig:map} with flux $>$~50 mCrab.  Sources were included only if the distance between the peak flux and the actual source position (final column) was $<$~3$^{\circ}$.}
    \renewcommand{\arraystretch}{1.2}
    \begin{tabular}[t]{lcccc}
      \hline
Source & Peak & Flux & Peak & Distance \\
Name   & $\sigma$ & (mCrab) &(l,b)&(degs)\\ \hline
        Crab &   212 &  1000& ( -177,   -6) &     1.58\\
      CygX-1 &   154 &   727 & (   71,    4) &     0.99\\
      ScoX-1 &    71 &   338 & (   -1,   24) &     0.24\\
      4U1700 &    67 &   319 & (  -13,    2) &     0.77\\
      1E1740 &    62 &   294 & (   -1,    0) &     1.05\\
     OAO1657 &    60 &   282 & (  -15,    2) &     1.44\\
     GX354+0 &    58 &   273 & (   -5,    0) &     0.72\\
     VelaX-1 &    49 &   231 & (  -99,    4) &     2.06\\
    4U1630-47 &   40 &   188 & (  -25,    0) &     1.93\\
     CygX-3 &    38 &   181 & (   77,    0) &     2.92\\
    GRS1915 &    28 &   132 & (   43,    0) &     1.94\\
      AqlX-1 &    20 &    94 & (   35,   -2) &     2.26\\
     GX301-2 &    20 &    93 & (  -61,    0) &     1.10\\
      Mrk501 &    18 &    84 & (   61,   40) &     2.82\\
     PKS2005 &    17 &    80 & (   11,  -30) &     0.81\\
     Geminga &    16 &    77 & ( -163,    2) &     2.94\\
      4U1608 &    16 &    77 & (  -29,   -2) &     1.15\\
        CenA &    14 &    67 & (  -49,   18) &     2.06\\
       3C273 &    16 &    74 & (  -69,   62) &     2.58\\
      HR1099 &    14 &    65 & ( -175,  -40) &     1.57\\
       \hline \\
      \end{tabular}
    \label{tab:sources}
  \end{center}
\end{table}

\end{document}